\documentclass[twocolumn,showpacs,preprintnumbers,amsmath,amssymb]{revtex4}

\usepackage{graphicx}
\usepackage{dcolumn}
\usepackage{bm}

\begin{document}

\title{Rank-based model for weighted network with hierarchical organization and disassortative mixing}
\author{Liang Tian}
\author{Da-Ning Shi}
\author{Chen-Ping Zhu}
\affiliation{College of Science, Nanjing University of Aeronautics
and Astronautics, Nanjing, 210016, PR China}

\date{\today}

\begin{abstract}
Motivated by a recently introduced network growth mechanism that
rely on the ranking of node prestige measures [S. Fortunato \emph{et
al}., Phys. Rev. Lett. \textbf{96}, 218701 (2006)], a rank-based
model for weighted network evolution is studied. The evolution rule
of the network is based on the ranking of node strength, which
couples the topological growth and the weight dynamics. Both
analytical solutions and numerical simulations show that the
generated networks possess scale-free distributions of degree,
strength, and weight in the whole region of the growth dynamics
parameter ($\alpha>0$). We also characterize the clustering and
correlation properties of this class of networks. It is showed that
at $\alpha=1$ a structural phase transition occurs, and for
$\alpha>1$ the generated network simultaneously exhibits
hierarchical organization and disassortative degree correlation,
which is consistent with a wide range of biological networks.
\end{abstract}

\pacs{89.75.-k, 89.75.Hc}

\maketitle

A major source of the recent surge of interest in complex networks
has been the discovery that a large number of real-world networks
have a power-law degree distributions, so called scale-free networks
[1-4]. Due to the peculiar structural features and the critical
dynamical processes taking place on them [5-8], there has been a
tremendous number of works modeling networks with scale-free
properties. The previous models of complex networks always
incorporate the preferential attachment [4], which may results in
scale-free properties. That is, a newly added node is connected to
preexisting one with a probability \emph{exactly} proportional to
the degree or strength of the target node. In reality, however, this
\emph{absolute} quantity information of an agent is often unknown,
while it is quite common to have a clear idea about the
\emph{relative} values of two agents. For this perspective,
Fortunato \emph{et al}. propose a criterion of network growth that
explicitly relies on the ranking of the nodes according to the
prestige measure [9]. This rank-based model can well mimic the
reality in many real cases that the \emph{relative} values of agents
is easier to access than their \emph{absolute} values. Motivated by
their work, we propose a model for weighted network evolution with
only ranking information available. Analytically and by simulations,
we demonstrate that the generated networks possess scale-free
distributions of degree, strength, and weight. The clustering and
correlation properties of this class of networks are also
investigated.

A weighted network is often denoted by a weighted adjacency matrix
with element $w_{ij}$ representing the weight on the link connecting
node $i$ and $j$. In the case of undirected graphs, weights are
symmetric $w_{ij}=w_{ji}$, as we will focus on. A natural
generalization of connectivity in the case of weighted networks is
the node strength defined as $s_i=\sum_{j\in\mathcal{V}(i)}w_{ij}$,
where the sum runs over the set $\mathcal{V}(i)$ (neighbors of node
$i$). This quantity is a natural measure of the importance or
centrality of a node in the network. As confirmed by measurement,
weighted complex network not only exhibits a scale-free degree
distribution $P(k)\sim k^{-\gamma}$ with
$2\leqslant\gamma\leqslant3$ [10,11], but also the pow-law strength
distribution $P(s)\sim s^{-\eta}$ [11] and weight distribution
$P(w)\sim w^{-\theta}$ [12]. Highly correlated with the degree, the
strength usually displays scale-free property $s\sim k^\beta$
[13,14].

\begin{figure}
\scalebox{0.8}[0.8]{\includegraphics{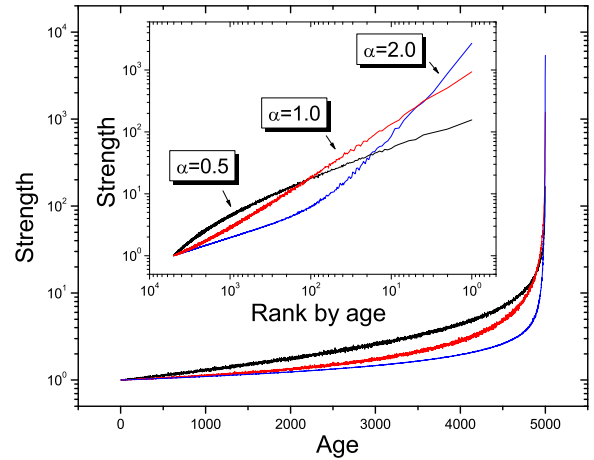}} \caption{(Color online)
Node strength versus node age with $\alpha=0.5$, $\alpha=1.0$, and
$\alpha=2.0$ from top to the bottom. Inset: Log-Log plot of the
relation between node strength and node rank by age. All the data
are averaged over $100$ independent runs of network size $N=5000$.}
\end{figure}

\begin{figure}
\scalebox{0.75}[0.9]{\includegraphics{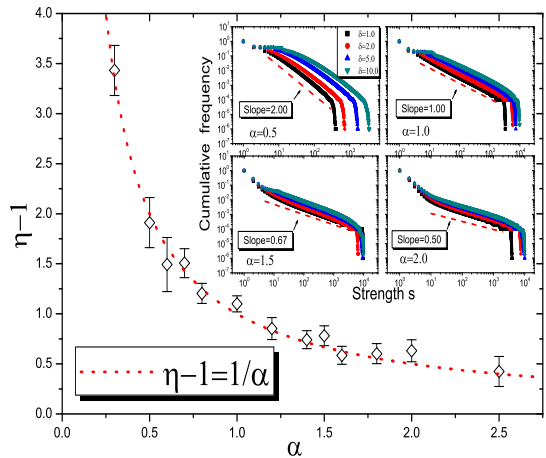}} \caption{(Color online)
The inset shows the cumulative strength distributions of networks
generated by using our model with different parameters $\alpha=0.5$,
$\alpha=1.0$, $\alpha=1.5$, and $\alpha=2.0$. The four dashed lines
have slopes 2.00, 1.00, 0.67, and 0.50 separately for comparisons.
The main plot shows the value of the strength distribution exponent
$\eta$ as a function of $\alpha$ obtained from numerical
simulations. The dotted line is the prediction of Eq. (6). All the
data are averaged over 100 independent runs of network size
$N=10^4$.}
\end{figure}

In the present model, the prestige ranking criterion is strength.
The definition of the model is based on two coupled mechanisms: the
topological growth and the weights' dynamics. The model dynamics
starts from an initial seed of $N_0$ nodes connected by links with
assigned weight $w_0$.

(1) \emph{Topological growth.} At each time step, a new node $n$ is
created and $m$ new links, with an assigned weight $w_0$ to each,
are set between node $n$ and pre-existing nodes. The previous nodes
are ranked according to their \textbf{strength}, and the linking
probability that the new node be connected to node $i$ depends on
the rank $R_i$ of $i$:
\begin{equation}\Pi_{n\rightarrow i}=\frac{R_i^{-\alpha}}{\sum_{\nu}R_\nu^{-\alpha}},\end{equation}
where $\alpha>0$ is a real-valued parameter. Note that the larger
the rank of the node is, the more difficult for it to gain new
links, which is reasonable in real life.

(2) \emph{weights' dynamics.} Analogous to the step in the model
proposed by Barrat \emph{et al}. (BBV model) [15], the introduction
of the new link on node $i$ will trigger local rearrangements of
weights on the existing neighbors $j\in\mathcal{V}(i)$, according to
the rule
\begin{equation}w_{ij}\rightarrow w_{ij}+\delta\frac{w_{ij}}{s_i},\end{equation}
where $\delta$ is the total induced increase in strength of node
$i$.

We firstly investigate the probability distribution of the generated
network. Since the strength-based ranking of a node can change over
time, it is hard to analyze the model directly by the ranking of
node strength. However, for a growing weighted network, there is a
strong correlation between the age of node and its strength, as the
older nodes have more chances to receive links. For these
considerations, we check this correlation by numerical simulations.
Fig. 1 shows the node strength as a function of its age. It can be
found that the function is monotone increasing with certain
fluctuations. Therefore, in the following \textbf{theoretical
analyse}, we make an approximation that we use the ranking by age
instead of that by strength. It will be showed by numerical
simulations that this approximation is reasonable.

The network growth starts from an initial seed of $N_0$ nodes, and
continues with the addition of one node per unit time, until a size
$N$ is reached. Hence, each node is labeled with respect to the time
step of its generation, and the natural time scale of the model
dynamics is the network size $N$. If the nodes are sorted by age,
from the oldest to the newest, the label of each node coincides with
its rank, i. e., $R_i=i \forall i$. Therefore, the node strength
$s_i$ is updated according to this evolution equation:
\begin{eqnarray}
  \frac{ds_i}{dt} &=& m\frac{R_i^{-\alpha}}{\sum_j R_j^{-\alpha}}(1+\delta)+\sum_{j\in \mathcal{V}(i)}m\frac{R_j^{-\alpha}}{\sum_l R_l^{-\alpha}}\delta\frac{w_{ij}}{s_j} \nonumber \\
   &=& m\frac{i^{-\alpha}}{\sum_j j^{-\alpha}}(1+\delta)+\sum_{j\in \mathcal{V}(i)}m\frac{j^{-\alpha}}{\sum_l
   l^{-\alpha}}\delta\frac{w_{ij}}{s_j}.
\end{eqnarray}
Using the continuous approximation, we treat $s$, $w$, and time $t$
as continuous variables and approximate the sums with integrals.
Solving Eq. (3) yields the strength evolution equation:
\begin{equation}
    s_i(t)\sim (\frac{t}{i})^{\alpha}
\end{equation}
Consequently, we can easily obtain in the infinite size limit the
probability distribution:
\begin{equation}
    P(s)\sim s^{-(1+1/\alpha)},
\end{equation}
which shows that the strength distribution of the network follows a
power law with exponent $\eta=1+1/\alpha$ for any value of $\alpha$.

Similarly to the previous quantities, it is possible to obtain
analytical expressions for the evolution of weights and the relative
statistical distribution. The rate equation of weight $w_{ij}$ can
be written as:
\begin{eqnarray}
  \frac{dw_{ij}}{dt} &=& m\frac{R_i^{-\alpha}}{\sum_j R_j^{-\alpha}}\delta\frac{w_{ij}}{s_i}+m\frac{R_j^{-\alpha}}{\sum_j R_j^{-\alpha}}\delta\frac{w_{ij}}{s_j} \nonumber \\
   &=& m\frac{i^{-\alpha}}{\sum_j j^{-\alpha}}\delta\frac{w_{ij}}{s_i}+m\frac{j^{-\alpha}}{\sum_j
   j^{-\alpha}}\delta\frac{w_{ij}}{s_j}.
\end{eqnarray}
Incorporating with Eq. (4), the above equation can be solved that
$w_{ij}\sim (t/t_{ij})^{2\delta(1-\alpha)}$, where $t_{ij}=max(i,j)$
is the time at which the edge is established. Therefore, the
probability distribution $P(w)$ is in this case also a power law
$P(w)\sim w^{-\theta}$, where
\begin{equation}
    \theta=1+\frac{1}{\alpha}+\frac{1}{2\alpha\delta}.
\end{equation}

In order to check the analytical predictions, we performed numerical
simulations of networks generated by the present model, where the
prestige ranking criterion is strength. In the inset of Fig. 2, we
plot the cumulative strength distributions of the networks
corresponding to various values of the exponent $\alpha$. In the
logarithmic scale of the plot, they exhibit power-law behaviors in
agreement with theoretical results. The relation between $\alpha$
and the exponent $\eta$ of the strength distribution is showed in
the main plot of Fig. 2, which confirms the validity of Eq. (6).
Together, the power-law distribution of weight $P(w)$ is shown in
Fig. 3. The analytical predictions can be perfectly confirmed by
numerical simulations. Noting the weights' dynamics step in the
definition of the model, the triggered increase $\delta$ is only
arranged locally. Therefore, we expect the proportionality relation
$s\sim k$, by which we easily obtain the scale-free distribution of
degree $P(k)\sim k^{-\gamma}$ with $\gamma=\eta=1+1/\alpha$. Since
there exist no new properties, we do not show them again here.

\begin{figure}
\scalebox{0.75}[0.8]{\includegraphics{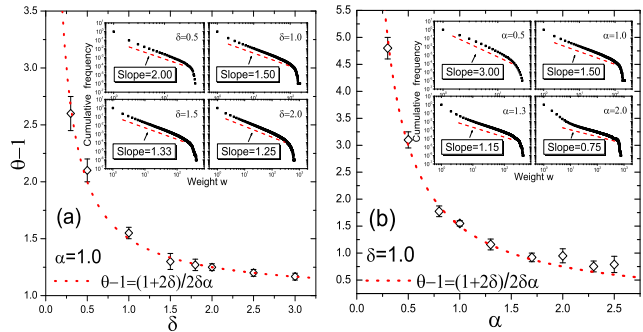}} \caption{(Color online)
Insets: Cumulative weight distributions of networks built according
to the present model for different value of (a) $\delta$ and (b)
$\alpha$. The main plots show the value of the weight distribution
exponent $\theta$ as a function of $\delta$ and $\alpha$ obtained
from numerical simulations. The dotted lines are the prediction of
Eq. (8). All the data are averaged over 100 independent runs of
network size $N=10^4$.}
\end{figure}

\begin{figure}
\scalebox{0.75}[0.8]{\includegraphics{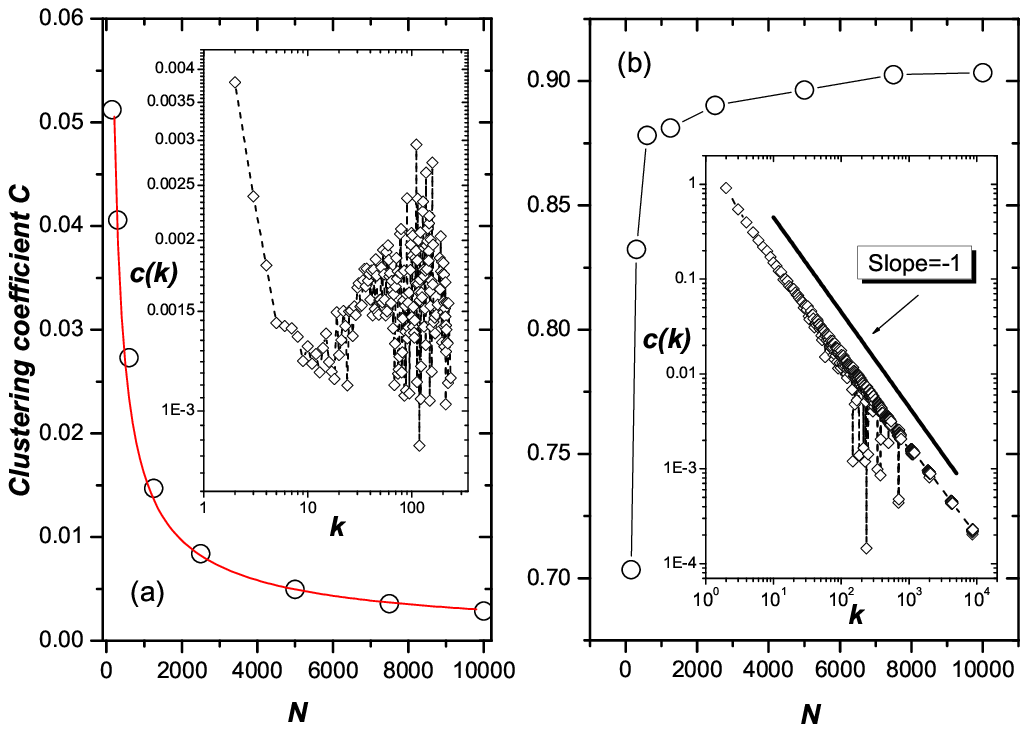}} \caption{(Color online)
Illustration of the average clustering coefficient $C$ as a function
of networks size $N$ for (a) $\alpha=0.5$ and (b) $\alpha=2.0$. The
insets show the behavior of $C(k)$ depending on degree $k$. The
curves in (a) is fit to algebraic decay form, $2.50\times
N^{-0.73}$. The solid line in the inset of (b) has slope -1 for
comparisons. All the data are averaged over 100 independent runs.}
\end{figure}

\begin{figure}
\scalebox{0.75}[0.8]{\includegraphics{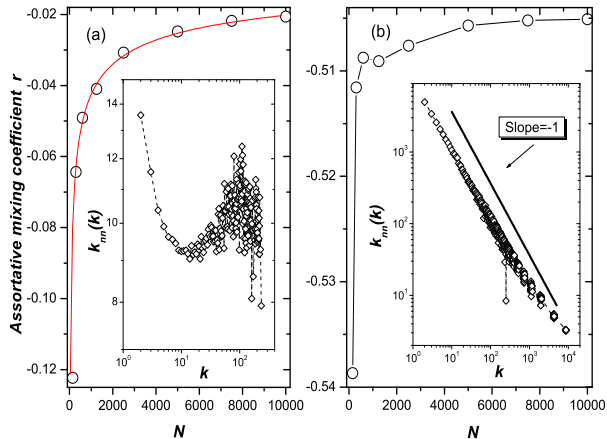}} \caption{(Color online)
Illustration of the assortative mixing coefficient $r$ as a function
of networks size $N$ for (a) $\alpha=0.5$ and (b) $\alpha=2.0$. The
insets show the behavior of $k_{nn}(k)$ depending on degree $k$. The
curves in (a) is fit to algebraic decay form, $-0.40\times
N^{-0.32}$. The solid line in the inset of (b) has slope -1 for
comparisons. All the data are averaged over 100 independent runs.}
\end{figure}

To better understand the topology of our model networks, we also
studied the clustering and correlation depending on the model
parameters $\alpha$. The clustering of a node $i$ is defined as [3]
\begin{eqnarray}C_i=\frac{2E_i}{k_i(k_i-1)},\end{eqnarray}
where $E_i$ is the number of links between neighbors of node $i$. It
measures the local cohesiveness of the network in the neighborhood
of the node. The average over all nodes gives the network clustering
coefficient $C$, which describes the statistics of the density of
connected triples. Further information can be gathered by inspecting
the average clustering coefficient $C(k)$, which denotes the
expected clustering coefficient of a node with $k$ degrees. In many
networks, the average clustering coefficient $C(k)$ exhibits a
highly nontrivial behavior with a power-law decay as a function of
$k$ characterizing the intrinsic hierarchy of the topology [16]. For
$\alpha=0.5$, the clustering coefficient $C$ seems to converge to
zero. This is seen by the accurate fits to algebraic decay forms in
Fig. 4 (a). Meanwhile, $C(k)$ is uncorrelated with $k$, denoting
that the network does not possess hierarchical structure. For
$\alpha=2.0$, $C$ approaches a stationary value of about 0.9 in the
limit of large $N$, which is showed in Fig. 4 (b). In this case, a
simple scaling form of clustering coefficient, $C(k)\sim k^{-1}$, is
obtained, which indicates that the network topology exhibits
hierarchical manner.

Another commonly studied network property is the degree correlation
of node $i$ and its neighbor. The average nearest neighbor degree of
node with connectivity $k$, $k_{nn}(k)$, is proposed to measure
these correlations. If degrees of the neighboring nodes are
uncorrelated, $k_{nn}(k)$ is a constant. When correlations are
present, two main classes of possible correlations have been
identified: \emph{assortative} behavior if $k_{nn}(k)$ increases
with $k$, which indicates that large degree nodes are preferentially
connected with other large degree nodes, and \emph{disassortative}
if $k_{nn}(k)$ decreases with $k$, which denotes that links are more
easily built between large degree nodes and small ones. A simpler
measure to quantify this structure is assortative mixing coefficient
[17]:
\begin{eqnarray}
  r=\frac{L^{-1}\sum_{i}j_ik_i-[L^{-1}\sum_i\frac{1}{2}(j_i+k_i)]^2}{L^{-1}\sum_{i}\frac{1}{2}(j_i^2+k_i^2)-[L^{-1}\sum_i\frac{1}{2}(j_i+k_i)]^2},
\end{eqnarray}
where $j_i$,$k_i$ are the degrees of nodes at the ends of the $i$th
edges, with $i=1,...,L$ ($L$ is the total number of edges in the
graph). This quantity takes values in the interval $[-1,1]$, where
positive values mean \emph{assortative} and negative values mean
\emph{disassortative}. Fig. 5 shows the simulation results. When
$\alpha=0.5$, the value of $r$ converges algebraically to zero, and
$k_{nn}(k)$ is unrelated with $k$, which denotes that correlations
are absent. On the contrary, when $\alpha=2.0$ the assortative
mixing coefficient is almost independent of network size for large
$N$. Meanwhile, $k_{nn}(k)\sim k^{-1}$, characterizing the
\emph{disassortative} degree correlation in the network.

To sum up, we studied a rank-based model for weighted network. The
scale-free properties of probability distributions of degree,
strength, and weight are obtained analytically and by simulation.
Furthermore, we investigate the clustering and correlation of the
network. It is indicated that a structural phase transition occurs
when the growth dynamics parameter $\alpha=1$ [18]. For $\alpha<1$
($\gamma>2$), $C(k)$ and $k_{nn}(k)$ are observed as a horizontal
line subject to fluctuations, and clustering coefficient $C$ and
assortative mixing coefficient $r$ converge to zero in the large
limit of network size $N$. For $\alpha>1$ ($1<\gamma<2$), there
emerge a few hub nodes in the network which are linked to almost
every other site ,and the generated network exhibits hierarchical
topology and \emph{disassortative} degree correlation. Moreover, the
clustering coefficient $C$ is independent of network size $N$ and
approaches a high value. Interestingly and specially, in this region
of parameter $\alpha$, the generated networks can well mimic the
biological networks which always appear to be \emph{disassortative}
[3,17] and possess hierarchical organization [16,19]. We think that
this class of network provides us with a new method to reconstruct
the hierarchies and organizational architecture of biological
networks, and it may be beneficial for future understanding or
characterizing the biological networks.

\section*{ACKNOWLEDGMENTS}

\end{document}